\documentstyle[12pt,aaspp4]{article}

\begin{document}

\title{Brown dwarfs in the Hyades and beyond?}
\footnote{Based partially on observations made with the Keck telescope}

\author {I. Neill Reid}
\affil {Palomar Observatory, 105-24, California Institute of Technology, Pasadena, CA 91125,
e-mail:  inr@astro.caltech.edu}

\author {Suzanne L. Hawley}
\affil {Michigan State University, East Lansing, Michigan, USA
email: slh@pillan.pa.msu.edu}

\begin{abstract}

We have used both the Low-Resolution Imaging Spectrograph and
the HIRES echelle spectrograph on the Keck telescopes to obtain 
spectra of twelve candidate members of the Hyades cluster identified 
by Leggett and Hawkins (1988, 1989). All of the objects are 
chromospherically-active, late-type M-dwarfs, with H$\alpha$
equivalent widths varying from 1 to 30\AA. Based on our measured radial
velocities, the level of stellar activity and other spectroscopic
features, only one of the twelve stars has properties consistent with
cluster membership. We consider how this result affects estimates of
the luminosity and mass function of the Hyades cluster.
Five of the eleven field stars have weak K I 7665/7699\AA \ 
and CaH absorption as compared with M-dwarf standards of the same spectral type, 
suggesting a lower surface gravity. Two of these sources, LH0416+14 
and LH0419+15, exhibit significant lithium 6708 \AA \ absorption.
Based partly on parallax measurements by the US Naval Observatory (Harris et al, 1998), 
we identify all five as likely to be young, pre-main sequence objects in or
near the Taurus-Auriga association at distances of between 150 and 250 parsecs.
A comparison with theoretical models of pre-main sequence stars indicates masses 
of less than 0.05 M$_\odot$. 

\end{abstract}


\section{Introduction}

Apart from the few stars in the insubstantial Ursa Major system, 
the Hyades cluster is the nearest open cluster to the Sun. The cluster centre
lies at a distance of $\sim46.3$ parsecs (Perryman et al, 1998), although
individual members are found at radial distances of 15 parsecs or more. The
cluster has a substantial space motion relative to the Sun, leading to transverse 
angular motions of from 0.09 to 0.19 arcsec yr$^{-1}$. As a result, successive
proper motion surveys (van Buren, 1953;  van Altena, 1969; 
Hanson, 1975; Luyten, Hill \& Morris, 1981; Schwan, 1990; Reid, 1992; Bryja, Humphrey \& Jones, 
1994) have succeeded in identifying cluster members with spectral types as late 
as M6 to M7, and absolute magnitudes, M$_V < 16.5$. Reid's (1992) deep proper-motion
survey and subsequent photometric follow-up observations demonstrate
conclusively that significant mass segregation is present within the cluster (Reid, 1993; 
Perryman et al, 1998). This is 
not unexpected, given an age of 625$\pm50$ Myrs (Perryman et al, 1998).
The concomitant preferential evaporation of low-mass stars means that
the Hyades is of only limited utility in studying matters concerned with
the stellar mass function. However, the cluster stars continue to play a
major role in calibrating stellar evolutionary models, in supplying a benchmark for
analysis of time variation in stellar activity, in probing stellar
binarity within a specific environment, and in providing an important
link in the extragalactic distance scale.

The lowest-mass Hyades stars are of considerable interest in studies of
evolutionary behaviour, particularly in comparison with results derived for
comparable stars in the $\approx$125 Myr-old Pleiades cluster. The latter
system is more distant, at $\sim$130 parsecs (Pinsonneault et al, 1998),
but paradoxically easier to survey, given both the higher intrinsic
luminosity and temperature of the younger low-mass stars (and brown dwarfs), and
the smaller solid angle subtended and consequent higher surface density
of cluster members.
Since the Hyades covers more than 200 square degrees, 48-inch Schmidt
plate material remains the most effective means of identifying low-luminosity
members, where field stars outnumber cluster members by factors of from 1000 to
10000. The earliest-epoch data for proper-motion surveys are
supplied therefore by the 103aO (blue) and 103aE (red) plates taken in the course of the
POSS I survey. Those plates have limiting magnitudes of $\sim$21 in the blue and 
$\sim20.5$ in the red. In the case of late-type M-dwarfs
the former corresponds to an effective limiting magnitude of V$\sim$19, 
or M$_V\sim$15.7 at the average distance of the Hyades. Even using the
red plates alone gives an effective advantage of only $\sim 1$ magnitude, 
since the red plexiglass
filter used in POSS I (and Luyten's subsequent 1962/63 LP survey) isolates
the $\lambda\lambda 5900-6500$ \AA \ region - i.e. the bluer half of the
R passband. 

In an effort to extend coverage to Hyades members at significantly
lower luminosities, Leggett \& Hawkins (1988, 1989: LH88 and LH89
respectively) undertook a purely photometric survey of 
a single Schmidt field ($\sim28$ square degrees) in the central regions of the cluster.
Their initial analysis was based on a series of 3 IIIaF+RG630 (R-band) and 3
IVN+RG715 (I-band) plates taken at the UK Schmidt telescope in 1985 and
1986, and scanned by the COSMOS automated microdensitometer at the Royal Observatory,
Edinburgh. Candidate cluster members were identified based on the (R-I) photographic
colours (the IIIaF+OG590 filter is a much closer match to the standard
R-band, Bessell, 1986), with subsequent JHK near-infrared photometry
used to eliminate obvious non-members. Many of the photometric
candidates have been confirmed subsequently as Hyads based on
either proper-motion (Reid, 1992; Bryja et al, 1994) or radial
velocity (Stauffer et al, 1994) criteria.

The limiting magnitude of COSMOS scans of the IIIaF UK Schmidt plates is 
only R$\sim20$ (LH89), while the latest-type M-dwarfs have photographic
(R-I) colours exceeding 2.5 magnitudes (Bessell, 1986). To allow for
this circumstance, Leggett and Hawkins (1989) extended their survey to
include objects with I$\le$18 magnitudes, but detected only on the IVN plates.
The majority of those sources have optical-to-infrared colours classifying
them as mid-type M-dwarfs at distances of 50 to 100 parsecs beyond the Hyades, 
but twelve faint (I $> 15.5$) stars, nine with no R-band data, were 
identified as having (I-J) photometric parallaxes consistent with their 
lying within the spatial confines of the cluster itself (LH89, Table 4).
Subsequent photometry by Leggett, Harris \& Dahn (1994) shows
that all twelve also have optical colours consistent with membership,
but does not exclude the possibility of their being foreground or
background field stars. 

If these stars are members of the Hyades, then several have luminosities
compatible with sub-stellar masses. One star (LH0418+13) is confirmed
spectroscopically as an M-dwarf (LH89), but the spectrum is
of insufficient resolution to determine cluster membership. Accurate
astrometric or radial velocity data provide the most rigorous means
of segregating field and cluster stars, but it is only recently that
it has proved possible to obtain such observations for these faint
stars. Harris et al (1998) have obtained accurate parallax and
proper motion measurements for three of the LH89 sources; this
paper presents new optical spectroscopy of all twelve stars.

We have used both the Low-Resolution Imaging Spectrograph and
the HIRES echelle spectrograph on the Keck telescopes to observe the
low-luminosity LH89 candidate Hyads. Those observations and the
derived spectral types are described in sections 2 and 3 of
this paper, while section 4 analyses the individual likelihood of
cluster membership, considering both our results and Harris et al's
astrometric data. Section 5 discusses the likely present-day 
luminosity and mass functions of the Hyades in light of these
results. Section 6 considers five LH89 stars which are likely 
to be low-mass, pre-main sequence objects in star-formation
regions behind the Hyades, and our results are summarised in section 7.

\section{Spectroscopic Observations}

\subsection {Low-resolution spectroscopy}

All of the candidate Hyades members (table 1) were observed spectroscopically
in January 1998 using the Low-Resolution Imaging Spectrograph (LRIS - Oke et al, 1995)
on the Keck II telescope. Nine stars (I-band only detections from LH88)
were observed by the authors on January 18 and 19 (UT), while data for the remaining three
stars were obtained by D. Kirkpatrick and C. Beichman on January 24. In the
former case, we used the 600 l/mm grating blazed at 7500\AA\ to give a 
dispersion of $\sim$1.3 \AA\ pix$^{-1}$, and wavelength coverage from 5750 to 8700\AA.
The latter observations employed the 400 l/mm grating (blazed at 8500\AA) and
cover the wavelength range 6000 to 10,000 \AA \ at 1.9 \AA\ pix$^{-1}$. In both cases, a
1-arcsecond slit was used, giving a resolution of 4.6 pixels at the detector. Exposure
times were 900 seconds with the 600 l/mm grating, and range from 300 to 600 seconds 
with the lower-resolution setup. 

Both series of observations were reduced and calibrated using standard procedures
in the IRAF software package. A 1-second dark exposure was used to remove
the bias level, and quartz-lamp flat-field exposures were used to normalise
the response of the detector. The individual stellar spectra were extracted
using the 'apextract' routine in IRAF, allowing for the slight curvature of
a point-source spectrum viewed through the LRIS optics. Each spectrum has
sufficient signal to noise that the extraction parameters (the 'trace' on
the chip) were self-derived, with sky-subtraction achieved using similarly-extracted
arrays adjacent to the stellar spectrum. 

The overall form of the wavelength calibration for the extracted spectra is 
defined using observations of a neon-argon arc lamp, with the one-dimensional 
arc spectrum extracted matching the stellar template. Arc lamp exposures
were obtained on each night. However, since LRIS is subject to significant 
flexure with changes in altitude
and azimuth, we have used the position of the night-sky lines in the individual
spectra to set the zeropoint of that dispersion relation. 

Our spectra of the Hyades candidates are set on an absolute flux
scale using observations of either subdwarf standards (Oke \& Gunn, 1983)
or white dwarfs from the reference set calibrated
by Hamuy et al (1994).  HD 19445 was adopted as the standard star for the January 
18 and 19 observations, while LTT 1020 served that function on January 24. All
 observations were made with an OG570 filter placed in the beam,
eliminating any contribution from second-order radiation - a substantially
more important consideration for the flux standards than 
the late-type M-dwarf programme objects. Figures 1 and 2 present the 
fully-calibrated spectra of the twelve candidate Hyades members.

\subsection{High-resolution spectroscopy}

In addition to our LRIS observations, we were able to obtain spectra of eleven
of the stars listed in Table 1 using HIRES (Vogt et al, 1994) on the Keck I
10-metre telescope. Eight stars were observed on July 25 and 26 and three stars on August 24 (UT), 
1998 with the specific aim of obtaining accurate radial velocities. (LH0420+15
lacks HIRES data.) The data cover the wavelength range $\sim6500 - 8950\AA$,
with gaps in wavelength coverage between orders, with the August data covering a
slightly different wavelength range. The spectra were
bias-subtracted, flat-fielded and extracted using a suite of programmes written
by Tom Barlow, with the wavelength calibration defined based on IRAF analysis of
a thorium-argon arc taken at the start of the night. HIRES resides on a Nasmyth
platform, and experience has shown that this procedure gives a wavelength
calibration which is stable to better than 1.5 kms$^{-1}$. During the current observations,
arcs taken at the beginning and end of each night agree to within 1.0 kms$^{-1}$,
which is more than adequate for the present purposes.

Exposure times were only 450 to 600 seconds, but still provide 
sufficient signal in each spectrum for either the detection
of H$\alpha$ emission and/or cross-correlation using the reddest orders ($\lambda > 8600$\AA).
The nearby star Gl 83.1 (V$_{rad} = -29.0$kms$^{-1}$, Delfosse et al, 1998), 
observed on each night, was used as a template for cross-correlation, and unambiguous
correlation peaks (amplitude $>0.3$) were obtained for seven stars. Based on the
agreement in V$_{rad}$ derived for the four reddest orders, we estimate
internal uncertainties of $\pm1.5$ kms$^{-1}$ and external uncertainties of $\pm2$ kms$^{-1}$.
Cross-correlation the two July observations of Gl 83.1, calibrated independently, shows an
agreement in velocity of 0.5 kms$^{-1}$.

The remaining four stars (LH0416+14, LH0418+15, LH0429+15 and LH0429+17) are clearly a poor match 
to the Gl 83.1 template, and we have determined velocities for these stars by measuring the
central wavelength of the H$\alpha$ emission line. We estimate that the measured centroid is
accurate to $\pm0.1\AA$, or $\pm5$ kms$^{-1}$. Our Gl83.1 observations provide a check
on the velocity zeropoint of this technique: in each of the three observations the 
H$\alpha$ centroid indicates a radial velocity within 0.3 kms$^{-1}$ of the 
known heliocentric velocity. Comparing H$\alpha$ measurements against 
cross-correlation velocities for five stars which correlate successfully 
against Gl 83.1 (LH0416+16 and LH0422+17 have barely detectable H$\alpha$ emission)
gives $\Delta V = V_\alpha - V_{CC} = 1\pm6$ kms$^{-1}$. While the H$\alpha$ line is stronger 
and better defined in the four stars noted above, to be conservative we assign an 
uncertainty of 6 kms$^{-1}$ to the radial velocities derived from H$\alpha$ alone.

\section {Spectral classification}

Our spectroscopic observations confirm that all twelve LH objects are late-type
M-dwarfs, as expected given the extremely red (R-I) and (I-K) colours. All have
H$\alpha$ emission, indicating various degrees of chromospheric activity. Table 1
lists the measured equivalent widths, accurate to $\pm0.2$\AA. We have
used several techniques to estimate spectral types for these stars, in each case
tying the measurements to the Kirkpatrick et al (1993) scale. 
Martin, Rebolo and Zapatero-Osorio (1996) have
devised and calibrated several narrowband pseudo-continuum colour indices, and
we have measured their PC2 index (F(7560\AA)/F(7040\AA)) for each star. The
PC3 index (F(8245\AA)/F(7560\AA)) is also available from our lower-resolution
LRIS spectra. Our calibration also employs both the TiO5 index, (F(7130\AA)/F(7044\AA), which 
measures the full depth of the $\gamma$ 7053\AA \ TiO bandhead (Reid, Hawley \&
Gizis, 1996), and the (I-K) colour index (described below). 
Finally, we also calculate Kirkpatrick 
et al's (1995) VO and CaH (F(7035)/F(6975) indices, although we use only
the former in our spectral classification. All of these indices are
listed in Table 1.

Figure 3 plots the spectral-type/colour correlations for TiO5 and (I-K), 
where the calibrators are 
stars from the 8-parsec sample classified by Henry et al (1995). 
The TiO5 measurements are from Reid, Hawley \& Gizis (1995),
while the photometry is from Leggett's (1992) compilation. Martin et al
and Kirkpatrick et al calibrate their indices onto the same system. 
Both TiO5 and the PC2
index are double-valued, with turning points at M7 and M8 respectively, but
the other indices (notably VO) allow resolution of any ambiguity. 
The classifications generally agree to within $\pm0.5$ of a spectral class
and Table 1 lists the average value for each star. 

\section {Cluster membership}

Confirming these stars as late-type M-dwarfs, even chromospherically-active
M-dwarfs, does not provide confirmation
of cluster membership. The local space density of M6 to M8 dwarfs is
$\sim 7 \times 10^{-3}$ stars pc$^{-3}$ (Reid \& Gizis, 1997a), 
while the effective sampling volume for
the colour-selected Hyades sample is $\approx 700$pc$^{3}$ (photometric
parallaxes indicating $20 < r <70$pc). Hence,
one expects $\approx5$ late-type field dwarfs in the present sample. We
can identify those interlopers using our radial velocity measurements
and the overall spectroscopic properties.

\subsection {Radial velocities}

The Hyades cluster has a well-defined space velocity (V = 46.6 kms$^{-1}$; Perryman
et al, 1998) and a low velocity dispersion ($\sigma \sim 0.25$kms$^{-1}$).
Hence we can predict the expected radial velocity for each LH star. Since all
of the stars lie within a relatively small area near the cluster centre, those
predicted velocities all lie in the range 37 {--} 39 kms$^{-1}$. Table 1 shows that only
one star, LH0418+13, has a measured radial velocity within 2$\sigma$ of the 
expected value for a Hyades cluster member. Providing partial confirmation of
these results, Harris et al (1998) have measured parallaxes and proper motions
for three stars (LH0416+14, LH0419+15 and LH0429+17), and none match
either the distance or tangential motion of the Hyades cluster. Thus, these measurements 
effectively eliminate eleven of the twelve LH stars as possible low-luminosity
members of the Hyades cluster. 

\subsection{ Linestrengths and bandstrengths}

While the radial velocity data rule out cluster membership for the majority
of the LH dwarfs, it is evident from inspection of Figures 1 and 2 that
several stars stand out as having distinct, correlated spectral properties.
The most obvious effect is variation in the equivalent width of the neutral
potassium resonance doublet at 7665 and 7699\AA. Those features have equivalent
widths of $\sim10$ and 5\AA\ in the dwarfs LH0416+16, LH0420+17 and LH0422+17, 
line-widths comparable with those observed in nearby late-type M-dwarfs such as 
VB8 and VB10. However, the lines are either absent or only weakly present in
the five stars with strongest H$\alpha$ emission: LH0416+14, LH0418+15, LH0419+15, 
LH0429+15 and LH0429+17. In addition, the $\lambda7666\AA$\ TiO system is noticeably more
prominent in the latter stars, while the Na I $\lambda$8183/8195 doublet is
significantly weaker in LH0424+15 than in LH0427+12. We shall refer to these
five stars as KI-weak dwarfs.

While the reduced KI linestrength is correlated with strong chromospheric emission,
this characteristic is unlikely to be {\sl caused} by that activity.
Weak emission is present in the cores of the potassium lines in most late-type
dwarfs, but at a much lower level than emission in either sodium or calcium. 
Hence, if the weak potassium absorption were due to emission filling in
the line profile, one would expect to see a comparable effect, and even emission, 
at both Na I $\lambda$8183/8195\AA\ doublet and the Ca II near-infrared triplet.
Our observation of LH0418+15 shows that the Na I doublet is weaker, but not
filled in, while there is no evidence for Ca II emission. Moreover,
a strictly relative comparison between LH0419+15
and LH0420+13, or LH0418+15 and LH0427+12, two pairs of stars with nearly identical
spectral types and H$\alpha$ emission equivalent widths, argues against chromospheric
activity as the source of the spectroscopic anomalies.

An alternative, and more viable, explanation is suggested by the behaviour of
the $\lambda6800$\AA\ CaH bands. CaH bandstrength depends primarily on 
temperature and surface gravity, with a weak dependence on metallicity (Mould, 1976)
-- the increased prominence of hydrides in late-type subdwarfs reflects the
decreasing strength of TiO absorption rather than an increase in MgH or CaH (Gizis, 1997).
We have used Kirkpatrick et al's (1991) index (F(7035)/F(6975)) to
measure CaH bandstrength in the LH stars, and Figure 4 compares those measurements
against data for spectral standards (taking the latter from Kirkpatrick et al, 1991).
A visual comparison of the KI-weak and normal LH dwarfs suggests a qualitative difference 
in the shape of the spectrum at $\lambda6800-7000$\AA. Figure 4 confirms that, indicating 
that the CaH bands are less pronounced amongst the five stars with weaker potassium absorption.

All of these characteristics -- reduced CaH bandstrength, weak or absent KI and NaI absorption,
and stronger TiO $\lambda7666$\AA\ -- are present in low surface-gravity M giants, 
rather than M dwarfs. The faint apparent magnitudes and
late spectral types rule out any question of these stars being giants, but
low-mass T Tauri stars, such as V410-XR3 (Luhman et al, 1998) or GG Tau (White et al,
in prep.), also have low gravities. There is no star formation region
along the direct line of sight toward the centre of the Hyades cluster, but, as
discussed further in section 6, there are a number of isolated, higher-mass T Tauri stars
within this area (Neuhauser et al, 1997). These KI-weak stars may be their lower-mass
counterparts.

\subsection {Lithium}

Further support for the likely pre-main sequence nature of these objects is
given by the detection of lithium in two of the KI-weak stars. 
Magazzu et al (1993) have highlighted the use of lithium 
detection as a means of identifying low-mass brown dwarfs, but the 
$\lambda 6708$\AA\ absorption feature
has long been known as a characteristic of $\sim10^6$-year old, low-mass T Tauri 
stars (Bonsack \& Greenstein, 1960). In the former case 
the presence of lithium indicates central temperatures
below $2.5\times10^6$K, and a low mass; in the latter, a youthful age and 
insufficient time to complete lithium depletion. 

Figure 5 plots the 6600-6800\AA\ region for LH0416+14, LH0419+15, LH0429+15
and LH0429+17 -- the four KI-weak stars observed at a resolution of 1.3\AA pix$^{-1}$.
In each case, the wavelength scale has
been adjusted to zero velocity, and the location of the Li I 6708\AA \ line
is marked. Two objects, LH0416+14 and LH0419+15, show evidence for absorption
at that wavelength, with equivalent widths of 0.3$\pm$0.1\AA\ and 0.5$\pm$0.1\AA\
respectively. In comparison, Luhman et al (1998) measure an 
equivalent width of 0.45\AA\ for the 6708\AA\ line in V410-XR3. Using the lithium-depletion
predictions made as part of the Burrows et al (1993, 1997) model
calculations, the presence of lithium
in stars of this spectral type points to an age of less than 125 Myrs.
We also plot data for LH0422+17, the only other star to
show marginal evidence for a detection, with EW=0.7$\pm$0.3\AA. The signal-to-noise
in the latter spectrum is only 15, as oppose to over 30 in the former two cases.
Further spectroscopy is required to solidify these detections.

There is no evidence for detectable lithium in the only remaining candidate Hyad,
LH0418+13. That result, however, is not surprising, since with an age of 625 Myrs,
even 0.065M$_\odot$ brown dwarfs are predicted to have depleted lithium 
fully (Burrows, priv. comm. based on the Burrows et al, 1997 models). The Hyades
lithium edge is expected to lie at M$_{bol} \sim 14.5$, or I$\sim19$, over 
a magnitude fainter than the faintest objects in the current sample.

\subsection {Conclusions}

In summary, only one star -- LH0418+13 -- amongst the Leggett/Hawkins sample
has properties consistent with Hyades membership. Of the eleven field stars, six are
spectroscopically similar to the average late-type dwarf in the Solar Neighbourhood.
The remaining five stars have characteristics which indicate lower surface
gravity and suggest their identification with pre-main sequence objects. The
presence of lithium in two of the latter objects implies ages of less than 125 Myrs.

\section {LH0418+13 -- the lowest-mass Hyades member?}

The reduction of the Leggett/Hawkins sample to a single  candidate Hyades
member obviously limits the conclusions one can draw. Nonetheless, it is useful to
compare the activity level of that star against other more massive cluster members.
Moreover, the absence of any other candidate cluster members of similar or lesser
luminosity sets significant constraints on the form of the cluster mass function.

\subsection {Chromospheric activity}

While equivalent width is a useful parameter for comparing chromospheric emission
among stars of similar spectral type, it does not give an accurate portrayal of the
relative activity amongst stars spanning a wide range of
luminosities (effective temperatures). The peak of the energy distribution moves
to longer wavelengths with decreasing temperature, leading to lower continuum
levels at 6560\AA. Thus, a 5\AA \ equivalent width emission line in an M0 dwarf
corresponds to significantly higher chromospheric activity than the same 
feature in an M7 star. As described by RHM, a more effective parameter for
tracking relative stellar activity is the fraction of the total bolometric
luminosity emitted in the H$\alpha$ line, $L_\alpha \over L_{bol}$. In the case
of the Hyades, the mean ratio is $\approx 10^{-3.9}$, with a dispersion in
individual properties of almost a factor of two. 

We have used Tinney et al's (1993) (BC$_K$, (I-K)) relation
to estimate a bolometric magnitude for LH0418+13, and, for an assumed distance modulus
of 3.33 mag.(Perryman et al, 1998), we derive M$_{bol}=12.78$. This is over half a 
magnitude fainter than the previously-known faintest cluster member, LH108 
at M$_{bol}\sim 12.2$ (Stauffer et al, 1995). The H$\alpha$ flux measured from our spectrum
is F$_\alpha$=$1.3 \times 10^{-16}$erg cm$^{-2}$ sec$^{-1}$ leading to a value of
${L_\alpha \over L_{bol}} = 1.7 \times 10^{-5}$. Figure 6 compares that activity level
against the (logL$_\alpha$/L$_{bol}$, M$_{bol}$) distribution of confirmed Hyades 
members, where the spectroscopic data are taken from Stauffer et al (1994), Reid, Hawley
\& Mateo (1995) and Leggett et al (1994). LH0418+13 lies at a significantly lower level than
the next faintest cluster members.

At face value, the relative inactivity of LH0418+13 seems to argue against cluster
membership. However, Stauffer et al (1995) have shown that in the Pleiades
cluster the L$_\alpha$/L$_{bol}$ ratio declines sharply at M$_{bol} \sim 11.5$,
corresponding to a mass of $\sim0.1 M_\odot$ in that younger cluster. We have
combined Burrows et al (1993) solar abundance, Y=0.25 (series X models) with the
more recent extension of those models to lower masses (Burrows et al, 1997)
to calibrate the Hyades-age (M$_{bol}$, mass) relation, and Figure 6 shows 
that M$_{bol} = 12$ corresponds to $\sim0.1 M_\odot$. The inferred mass for 
LH0418+13 is 0.083M$_\odot$, placing it very close to the hydrogen-burning limit.
Thus, while LH0418+13 may yet prove to be a field star, it remains possible that 
the reduced level of activity may echo the (as-yet unexplained) same phenomenon 
observed over the same mass range in the younger Pleiades cluster.

\subsection {The luminosity function}

The addition of stars from the LH survey marks a considerable extension to the range of
luminosity spanned by Hyades stars. We have calculated a revised luminosity 
function for the cluster by adding data for LH0418+13 to a 
reference sample derived by combining data from several prior Hyades surveys.
Our baseline sample consists of known Hyades members falling within the
112 square-degree region (four Schmidt fields) covered by Reid's (1992)
proper motion survey. Hyades members have been identified over a larger
area, primarily from Luyten's extensive proper-motion surveys, but the
more extended samples are less complete at faint magnitudes (Reid, 1993),
and follow-up observations by Leggett et al (1994) and others have
shown many low-probability candidates to be field stars.

Photographic saturation limits the proper motion
accuracy of Reid's survey at V$<10$ mag., but earlier surveys
by van Buren (1952) and Pels et al (1975) provide an accurate accounting at
bright magnitudes. The completeness limit at faint magnitudes is set by the
requirement that a star be detected on both the POSS I E and O plates to
determine the first epoch position. Matching Reid's survey against deeper studies, it
is evident that incompleteness sets in at m$_r \sim 18.5$, or V$\sim 19$. 
All of the stars within these limits have accurate photoelectric or CCD
photometry (Reid, 1993 and refs within; Leggett et al, 1994), while spectroscopy
is available for most stars, including all M-dwarfs.
Two hundred and thirty-three systems meet the astrometric, photometric and
spectroscopic criteria for cluster membership.

This dataset includes a number of known binaries. One hundred and twenty-eight
systems have high spatial resolution ($\sim 0.1$ arcsecond)
observations, either using the Hubble Space Telescope
Planetary Camera (Reid \& Gizis, 1997b) or infrared speckle imaging on the Hale
200-inch (Patience et al, 1998), and twenty-two are resolved as binaries. 
Only ninety-six systems have been monitored for radial velocity
variations, mainly by Griffin et al (1988), and thirty-nine of those stars are identified 
as binaries, including ten speckle binaries. In four of the latter cases, the spectroscopic
and speckle secondaries are identical. Lacking flux ratio data for most of the
unresolved spectroscopic binaries, we have not attempted to include those secondary 
components in our sample. However, we include in our reference sample data 
for resolved primary and secondary components. The overall impact of binarity can be 
gauged by comparing luminosity and mass function based on that sample against results 
derived from composite photometry.

Two studies extend to significantly fainter magnitudes: Bryja et al's proper-motion
survey; and the Leggett/Hawkins photometric survey. Both cover a single Schmidt field.
Bryja et al matched scans of the POSS I E plates against Luyten's 1963 E-plates
and a third-epoch E-plate taken in 1975. They 
therefore have an effective limiting magnitude of m$_r \sim 19.5$, or V$\sim 20$.
Their sample includes two confirmed Hyades members with V$>19$ (Br 262 = LP415-20,
and Br804 - Stauffer et al, 1994). Leggett and Hawkins' photometrically-selected
sample, excluding the twelve stars listed in Table 1, includes two 
candidates with R magnitudes fainter than 17, nos. 241 and 91 in LH87.
The former is RHy 138 (LP414-2008) in Reid (1992), initially suspected as spurious,
but later confirmed as a member (RHM; Stauffer, Liebert \& Giampapa (1995)). 
Leggett et al (1994) question whether LH91 is a cluster member based on its position in the
colour-magnitude diagram. We have, however, included both stars in our cluster sample,
assuming M$_I = 11.81$ for LH91. We also calculate bolometric luminosities using
the relation derived by Reid \& Gizis (1997a):
$$BC_I \qquad = \qquad 0.023 \quad + \quad 0.575 (V-I) \quad - \quad 0.155 (V-I)^2$$

We have combined these datasets by weighting each star based on the
solid angle covered by the parent survey: single weight for stars with V$<19$; 
quadruple weight for the Bryja et al dwarfs, LH 241, LH91 and LH0418+13. 
Absolute magnitudes are based either on the individual motions, scaling the 
mean cluster distance to match the Hipparcos-based modulus of 3.33 magnitudes
(rather than 3.40 magnitudes derived by Schwann, 1986), or are derived for an
assumed distance modulus of 3.33 magnitudes. In the case of brighter Hyades stars with 
only BV data, we use the (B-V)/(V-I) relation derived by Reid \& Gilmore (1982) to
estimate (V-I), and hence M$_I$. Figure 7 plots the derived
luminosity function. Allowing for the resolved binaries has little effect on the
results, which show a strong peak at M$_I = 8.75\pm0.25$ and
a subsequent sharp decline. 

\subsection {The mass function}

Since the Hyades cluster is both more metal-rich and younger than the
typical star in the Solar Neighbourhood, the luminosity function cannot be
compared directly with data for the local stars.  Moreover, there is clear
evidence of mass segregation in the cluster, with M-dwarfs having a
substantially larger core radius than the more massive solar-type
stars (Reid, 1992; Perryman et al, 1998). As a result, one expects tidal effects to have led
to preferential evaporation of lower-mass cluster members. The relatively
low M-dwarf:G-dwarf number ratio in the cluster, as compared with
the field, tends to support that hypothesis - assuming similar initial mass functions.
Nonetheless, the present-day mass function (PDMF) of cluster stars remains of interest, 
particularly near the hydrogen-burning limit. Objects just above 
and just below the stellar mass limit should experience relatively
little differential dynamical evolution. Hence the PDMF provides
an indication of the form of the initial mass function, $\Psi(M)$,
over a limited mass range. 

With an age of 625 Myrs, low-mass stars in the Hyades are expected to have
higher luminosities than their older counterparts in the field. Burrows et
al (1993) have computed models for Hyades-age dwarfs with masses below 0.2 M$_\odot$, 
and those models (their series X) indicate a
negligible difference in luminosity between 600 Myr and 10 Gyr-old stars 
with masses above 0.125 M$_\odot$ (M$_{bol} < 11.4$).  The
mass-luminosity relation predicted by those models for the 0.125 to 0.2 M$_\odot$ mass
range is in good agreement with the empirical mass-luminosity 
relation derived for field dwarfs by Henry \& McCarthy (1993). We have
therefore combined the latter empirical calibration with the Burrows et al
X-models predictions to derive the present-day mass function of the Hyades.
For completeness, we extend our calibration to the higher-mass Hyades
stars, using the Henry \& McCarthy (mass, M$_K$) or (mass, M$_V$) relations
as appropriate. 

Figure 8 plots our derivation of the present-day mass function of the Hyades. Again, we
give quadruple weight to the lower-luminosity Bryja et al and Leggett/Hawkins dwarfs. 
The field mass function is well-represented by a power-law, $\Psi(M) \propto M^{-1.05}$
(Reid \& Gizis, 1997b), or a flat distribution in dlogN/dlogM. The Hyades
mass function plotted in Figure 8 has a slope of $\sim0.3$ between 0.25 and 1 M$_\odot$,
i.e. a less-steeply increasing mass function than the field, $\Psi(M) \propto M^{-0.7}$,
as one might expect given dynamical evolution and stellar evaporation. This is generally consistent
with Leggett et al's (1994) conclusion that power-law indices of $0 < \alpha < 1$ 
provide the best match to the Hyades I-band luminosity function. The
observed mass function, however, clearly falls below the extrapolation of that
relation at masses below 0.25 M$_\odot$, reflecting the corresponding dip in
$\Phi$(M$_I$) at M$_I$=10.5 in Figure 7. 

It is unlikely that the downturn in $\Psi(M)$ at 0.25 M$_\odot$ is due
to {\sl relative} incompleteness in the Hyades star sample. Although mass
segregation is clearly present, 
Raboud and Mermilliod (1998) find that in Praesepe, a cluster almost identical
in age to the Hyades, eighty percent of the cluster stars, regardless of mass,
lie within a radius of 1.5 degrees of the cluster centre. That limit corresponds to
a radius of 6 degrees at the Hyades, well within the region covered by Reid's
112-square degree survey which is complete to M$_I \sim 11.5$ ($\sim 0.11 M_\odot$).
Hence, Figure 8 should be representative of the overall 
stellar mass distribution in the cluster. The implication is that the present-day 
scarcity of M-dwarfs is due either to significantly more rapid evaporation at masses below
$\sim0.2 M_\odot$, or to an initial deficit.

Turning to the mass function near the substellar boundary, there are only
nine stars in the full sample with masses estimated as below 0.12 M$_\odot$, 
with LH0418+13 the only candidate for a mass lower than 0.1 M$_\odot$. 
Based on the magnitudes listed in Table 1, we can estimate an approximate
limiting I magnitude of 17.25 to 17.5 for the Leggett/Hawkins survey - not unreasonable
for UKST IVN photographic plates. This corresponds to an absolute magnitude limit
of M$_I\sim14$ for Hyades stars, or M$_{bol}\sim13.5$. Applying our adopted
mass-luminosity relation, the resulting mass limit is $\sim0.07 M_\odot$. Extrapolating 
the M$^{-0.7}$ mass function plotted in Figure 9, we would expect $\sim6$
Hyades stars with masses less than 0.1M$_\odot$ to have been detected
by Leggett \& Hawkins. In fact, we have only one candidate.

Leggett et al (1994) estimate the LH88/LH89 survey to be complete only at the 74\%
level, but that level of incompleteness can scarcely account for the almost
complete absence of
very low-mass members. It is worth re-emphasising that the Schmidt RI survey {\sl was}
successful in detecting late-type M-dwarfs with photometric properties
consistent with cluster membership -- it would be a remarkable coincidence
if (adopting the completeness estimate) all four of the hypothetical
M-dwarfs missed in their survey were Hyades members. Can binaries supply
the missing low mass stars? That seems unlikely, given the low fraction 
of binaries, and the relative scarcity of very low-mass companions, detected 
amongst M-dwarfs by Reid \& Gizis (1997b).
Moreover, Zapatero-Osorio et al (1997) derive an M$^{-1}$ mass function for the 
Pleiades based on isolated objects, with no corrections included for 
possible unresolved binaries.

Setting aside observational conspiracies, 
our analysis suggests that the Hyades mass function flattens significantly (fewer
low-mass objects) near the hydrogen-burning limit, and may even come 
close to truncation. If such proves to be the case, then either very low-mass stars and
brown dwarfs are ejected from clusters with high efficiency between ages of
125 and 625 Myrs, or the Hyades and Pleiades had different initial mass functions.
Dynamical simulations of open clusters (e.g. Fuente Marcos, 1995) do
predict preferential evaporation of lower-mass stars, but it is not clear
whether they can account for the almost total absence of objects near
the hydrogen-burning limit.

\section {The pre-main sequence stars}

In section 3 we identified a group of five stars from Table 1 which have
weak CaH and KI and which are probably low-mass
pre-main sequence (PMS) stars with lower gravity than comparable field and Hyades
stars.  Table 2 lists the basic data for these five stars, including parallax-based
distance  measurements from the US Naval Observatory (Harris et al, 1998) for three
stars. Those parallaxes place all three objects well above the colour-magnitude relation
defined by nearby dwarfs, as expected for protostars still contracting onto the 
main sequence.
The identification of low-mass PMS stars in a search for brown  dwarf members of a
nearby open cluster is not unprecedented - Oppenheimer et al (1997) showed that two
stars identified as Pleiades members by Hambly, Hawkins and Jameson (1993), HHJ 339 and 430,
are probably 20-Myr old PMS stars. The current detections are somewhat
more surprising, since there are no obvious star-forming regions within
the area covered by the Leggett/Hawkins (LH) survey field, centred at $4^h 24^m, +15^o$ 
(the Pleiades lies 20-parsecs in front of the Taurus-Auriga star-forming regions). However, 
Ungerecht \& Thaddeus' (1987) $^{12}$CO survey of the Taurus-Auriga complex identified
a small cloud lying at $4^h 27^m, +18^o$, on the north-eastern edge of the
field, and the IRAS 100$\mu m$ map constructed by Neuhauser et al (1997) showed
extensive far-infrared emission over the entire LH field. The latter
study used spectroscopic follow-up of ROSAT sources to identify at least 33
early-type (FGK) T Tauri stars within a 300 square-degree region south of Taurus. 
Several of those sources, as well as a number of stars with weak lithium absorption, 
lie in the vicinity of the LH field. The typical radial velocities of those
stars lie between 10 and 20 kms$^{-1}$, consistent with the measurements
listed in Table 1. More significantly, many lie several degrees 
south of the nearest detected 100$\mu m$ emission, and over 20 degrees from the 
Taurus-Auriga clouds, despite having ages estimated at only $\sim$10 Myrs. 

The line of sight towards the LH field intersects part
of the Taurus-Auriga complex at a distance of $\sim 150$ parsecs from the Sun, 
a distance matching the parallax measurements for LH0429+17.
LH0416+14 and LH0419+15 have parallax measurements indicating somewhat
larger distances, and it is possible that these objects are part of 
an extension of the more distant Orion complex. To be conservative, we assume that the
remaining two PMS stars in the LH group share these larger distances. 
Table 2 lists the resulting inferred absolute I-band and bolometric magnitudes, 
using Tinney et al's (1993) (BC$_K$, (I-K)) relation to derive the latter.
While we believe these to be pre-main sequence stars, it is unlikely 
that they are sufficiently young that substantial
residual material remains in a disk. In support of that hypothesis, Luhman et 
al (1998) find no evidence for excess emission from V410-XR3 in either the V 
or L' passbands. Thus, the main sequence-calibrated 
bolometric corrections should be reliable to $\pm 0.1$ magnitude.
Given the low level of 100$\mu m$ emission 
within the field, these are uncorrected for any foreground reddening. 

In principle, one can estimate masses for PMS stars by comparing their location on the
HR diagram with the predictions of theoretical models. The main complication lies
in establishing a reliable temperature scale. Effective temperatures for
late-type main-sequence dwarfs are still defined insecurely, particularly for
spectral types later than M5 (Leggett et al, 1996), but it is clear that
M giants of similar spectral type have significantly higher temperatures 
(Bessell et al, 1989).
Since pre-main sequence stars have lower gravities, and therefore more extended atmospheres
than main-sequence dwarfs, most recent studies of late-type PMS stars adopt 
temperature values
intermediate between the two scales, generally lying closer to the dwarf scale. 

Leggett et al (1996) provide the most recent study of the M-dwarf effective
temperature scale, combining extensive optical and near-infrared photometry and
spectrophotometry with model atmosphere analyses. We follow Luhman et al (1997, 
1998) in adopting this as our reference dwarf scale. One should note, however, that
the lowest luminosity star included in the Leggett et al study is GJ 1111, type M6.5 
(T$_{eff} \approx 2700$K). Hence, we are forced to extrapolate this scale
to lower temperatures to accommodate the LH stars. Those extrapolations are
guided partly by the offset between the Leggett et al and Kirkpatrick et al (1993)
scales, and partly by Luhman et al's (1997) estimate of T$_{eff} = 2500K$ for an
M8.5 main-sequence dwarf. 

Table 2 lists the dwarf-scale effective temperature estimates for the five
Leggett/Hawkins PMS stars. We also show the corresponding giant-scale
temperatures (T$^g$), based on the ((I-K), T$_{eff}$) calibration given
by Bessell et al (1989). Citing as-yet unpublished models by Baraffe, 
Luhman et al (1997, 1998) estimate that the appropriate temperatures for
PMS dwarfs are $\sim100$ to 200K hotter than the dwarf scale. Lacking any
empirical determinations for such low-luminosity objects, we have followed
suit in arriving at the adopted temperatures listed in Table 2.

Figure 9 compares the location of the five LH PMS stars in the HR diagram
against the evolutionary tracks computed by D'Antona \& Mazzitelli (1997).
We have also plotted data for the M6 PMS dwarf V410-XR3 (M$_{bol} = 7.6$, 
T$\sim 3040$K: Luhman et al, 1998) and for $\rho$ Oph 162349.8-242601
(M8.5, M$_{bol} = 11.1$, T$\sim2600$K; Luhman et al, 1997). All five of
the LH stars lie well to the right of the transition to the substellar r\'egime,
implying masses of $\approx 0.06 M_\odot$ or less, and ages of only $\sim1$ Myr.
LH0429+17 appears to have properties which are very similar to those of
$\rho$ Oph 162349.8-242601, save that the latter has more substantial H$\alpha$ emission
(EW$\sim 60\AA$). The exact location of all of these PMS objects, including V410-XR3 and the 
$\rho$ Oph source, depends critically on the temperature scale adopted. However, it appears 
highly likely that Leggett \& Hawkins succeeded in identifying brown dwarfs - at
distances of over 100 parsecs beyond the targeted Hyades cluster! 

Finally, we can compare the level of chromospheric activity in these low-mass
objects against that found in higher-mass stars of similar age. Hawley et al (1996) present
H$\alpha$ measurements for 26 members of the 30-Myr old cluster IC 2602. Figure 10 plots
those data, together with V410-XR3 and the two Pleiades interlopers identified
by Oppenheimer et al. Even allowing for the uncertainties in the bolometric magnitudes,
it is clear that all of these objects emit a substantially lower fraction of their
total luminosity at H$\alpha$ than the young cluster members. Presumably this
reflects the larger surface area, and hence higher photospheric flux of
these PMS stars, suggesting that they are indeed younger
than the IC 2602 stars.  In fact, observations of low-mass Pleiades stars 
suggest that if the LH dwarfs have masses below 0.1 M$_\odot$, they may never 
reach the activity level achieved by higher mass stars.

\section {Summary and Conclusions}

We have obtained and analysed intermediate-resolution spectra of twelve
red stellar objects identified as candidate Hyades members by Leggett \&
Hawkins (1988). It is clear that that survey achieved its goal of identifying
stars with colours consistent with cluster membership.   Our spectra show that 
all twelve stars are late-type, chromospherically-active M-dwarfs. However,
 radial velocity measurements made using the HIRES echelle indicate that at 
least eleven of these stars are not cluster members.

Five of the field M-dwarfs are identified as background pre-main sequence objects.
All have strong H$\alpha$ emission (${L_\alpha \over L_{bol}} > 10^{-4.2}$),
as well as relatively weak CaH and low K I 7665/7699 and Na I 8183/8195 \AA \
equivalent widths. Those features are characteristic of a lower surface gravity
than is present in main-sequence stars. Moreover, at least two of these low-gravity, 
active M-dwarfs have detectable Li 6708\AA \ absorption, while US Naval Observatory parallaxes
and proper motions show that three lie at distances matching the Taurus-Auriga
star-forming complex. Matching their locations on
the HR diagram against theoretical isochrones suggests that all five have 
masses below the hydrogen-burning limit, although the exact values 
are dependent on the adopted temperature scale.  The isochrones
indicate ages of only $\sim1$ Myr for these objects.

Of the twelve LH stars, only LH0418+13 has properties
consistent with Hyades membership. Adding that star to a complete sample
of known members, we have re-computed the luminosity function
and mass function for the cluster, and, making due allowance for
the effects of mass segregation,  find evidence for a significant
flattening in $\Psi(M)$ below $\sim0.25 M_\odot$. The scarcity of low-mass stars in 
the present-day mass function suggests either an efficient ejection
mechanism for those objects, or that the Hyades had a different
initial mass function than the Pleiades.

The Leggett \& Hawkins survey has proven to be extremely useful in
identifying interesting low-mass objects in the Hyades and
beyond.  Extending that survey to I$\sim 19$ would set strong
limits on the form of the Hyades mass function below the substellar limit,
and, perhaps, allow further detections of even lower-mass PMS stars 
in the background star-forming regions.

\subsection*{Acknowledgements}

We would like to thank Davy Kirkpatrick and Chas Beichman for obtaining
spectra of three of the M-dwarfs discussed in this paper, and Adam Burrows for 
making available the lithium-depletion predictions of his 1993 and 1997 stellar
models. The Keck Observatory
is operated by the Californian Association for Research in Astronomy, and
was made possible by generous grants from the Keck W. M. Foundation.  SLH
was partly supported by NSF (NYI) grant AST94-57455.

\clearpage
\centerline{FIGURE CAPTIONS}
\vskip2em

\figcaption{a-c: LRIS spectra at 1.3 \AA pix$^{-1}$ of nine LH stars}

\figcaption{LRIS spectra at 1.9 \AA pix$^{-1}$ of three LH stars}

\figcaption {Spectral-type/colour calibrations for the TiO5 and
(I-K) colour indices. The calibrating stars are drawn from the
northern 8-parsec sample, with TiO5 data from Reid et al (1995) and 
(I-K) photometry from Leggett (1992). }

\figcaption{CaH bandstrength. The crosses are spectral-type standards
from Kirkpatrick et al (1991); the open squares are KI-strong LH dwarfs; 
and the solid points are KI-weak LH stars.}

\figcaption{The $\lambda6600-6800$\AA\ spectral region in the four KI-weak
LH dwarfs stars observed at 1.3\AA\ pix$^{-1}$ resolution and the normal
dwarf LH0422+17. The vertical line indicates the expected position of 
the Li I $\lambda6708$\AA\ line.}

\figcaption{H$\alpha$ activity ratios for members of the Hyades cluster. 
The open triangle marks the location of LH0418+13, while the masses listed along the 
upper margin are based on predictions of the Burrows et al (1993, 1997) models.}

\figcaption{The I-band luminosity function of the Hyades cluster. The histogram
plots the results based on the systemic photometry (i.e. no allowance for binarity); 
the solid points, with Poissonian errorbars, show the effect of including 
explicitly the resolved binary components.}

\figcaption{The present-day mass function of the Hyades cluster. As in Figure 8,
the dotted histogram is the systemic mass function; the solid histogram includes
data for primary and secondary stars in resolved systems. The dashed line plots
the power-law mass function $\Psi(M) \propto M^{-0.7}$, which is an adequate fit
to the 0.25 to 1.0 M$_\odot$ mass range. The limiting magnitude of the
Leggett/Hawkins survey correspond to $\sim0.07 M_\odot$.}

\figcaption{The inferred locations of the LH PMS dwarfs in the HR diagram. The
theoretical isochrones and evolutionary tracks are from D'Antona \& Mazzitelli (1998),
with the hatched line marking the substellar boundary. The solid triangles mark
the locations of two known M-type PMS dwarfs, V410-XR3 and $\rho$ Oph 162349.8-242601
(Luhman et al, 1997; 1998).}

\figcaption{Chromospheric activity in pre-main sequence stars. The solid
squares are data for members of the 30-Myr old cluster IC 2602; the open circles
are our observations of the LH stars; the open triangles mark the location of V410-XR3
and the two Pleiades-field PMS dwarfs, HHJ 339 and 430.}

\clearpage
\setcounter {figure} {0}
\begin{figure}
\plotone{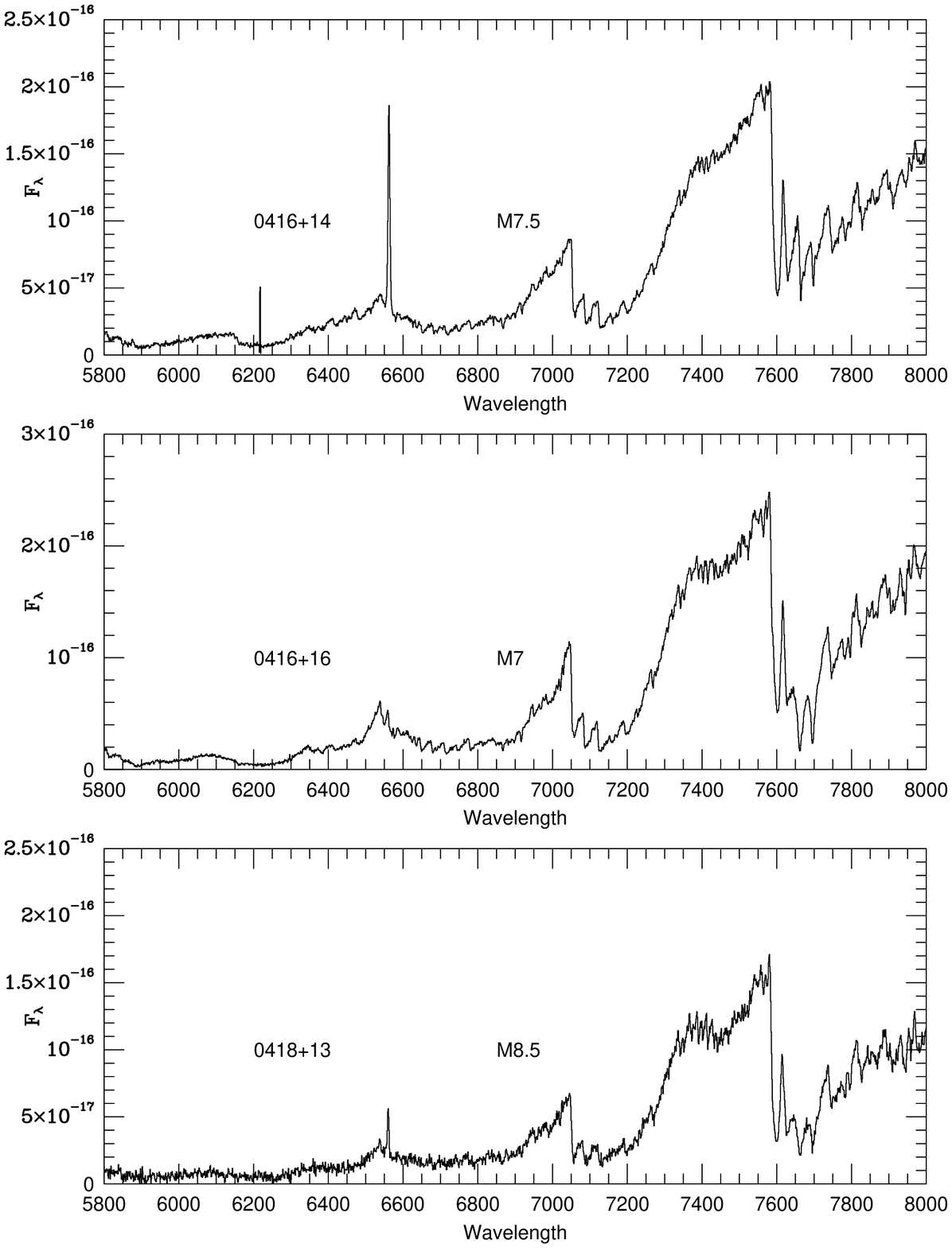}
\caption{ A: LRIS spectra at spectra at 1.3 \AA pix$^{-1}$}
\end{figure}

\setcounter {figure} {0}
\begin{figure}
\plotone{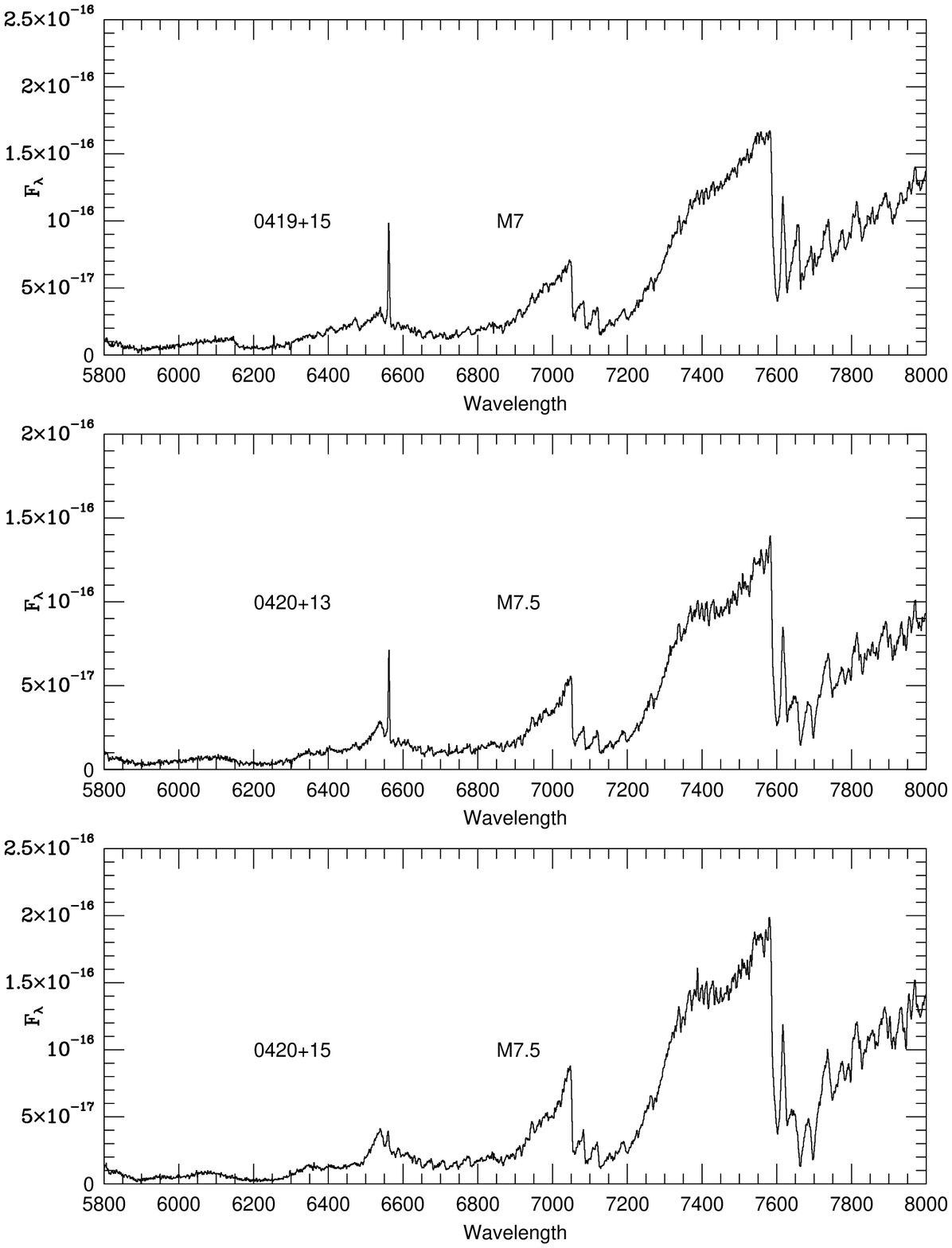}
\caption{ B: LRIS spectra at spectra at 1.3 \AA pix$^{-1}$}
\end{figure}

\setcounter {figure} {0}
\begin{figure}
\plotone{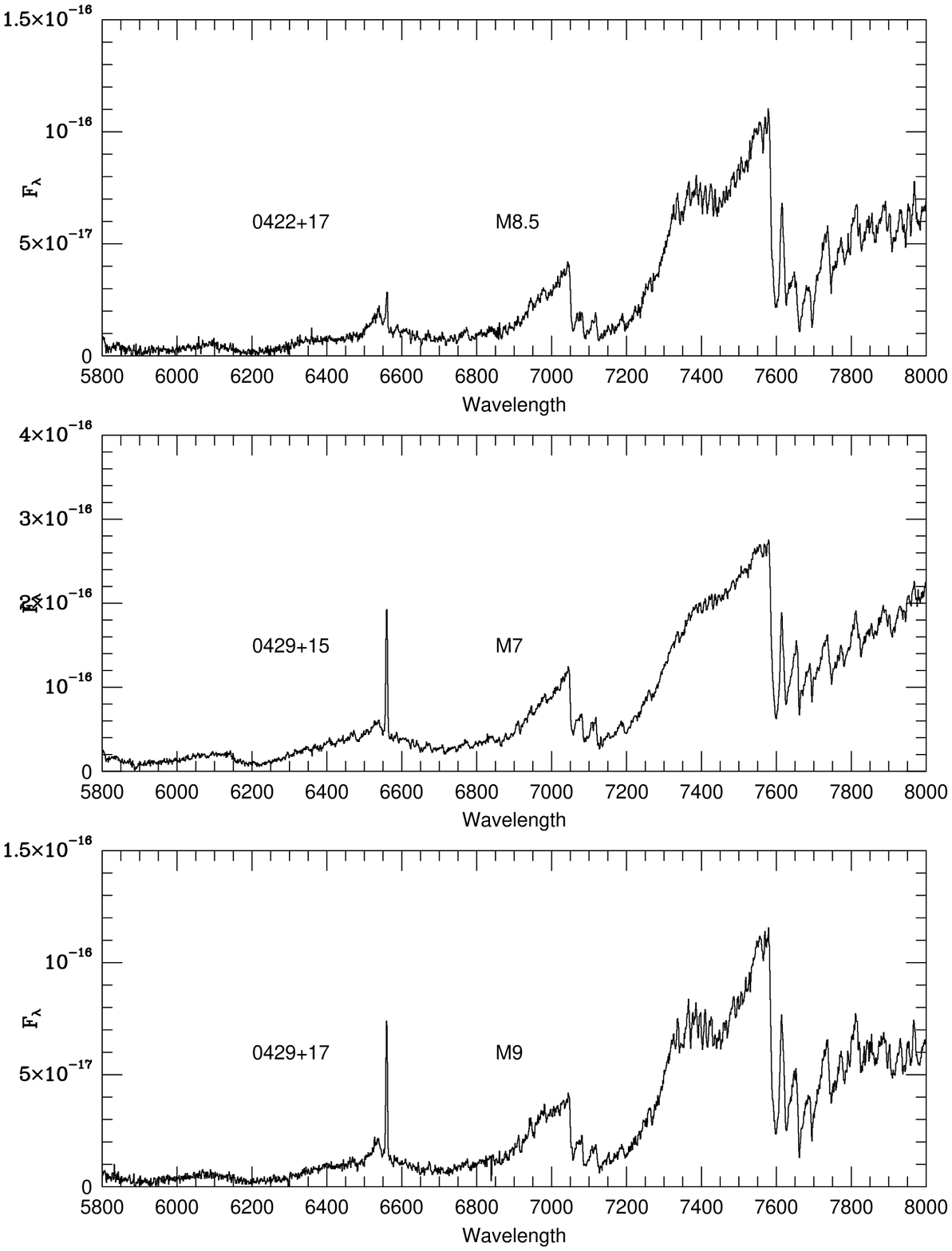}
\caption{ C: LRIS spectra at spectra at 1.3 \AA pix$^{-1}$}
\end{figure}

\begin{figure}
\plotone{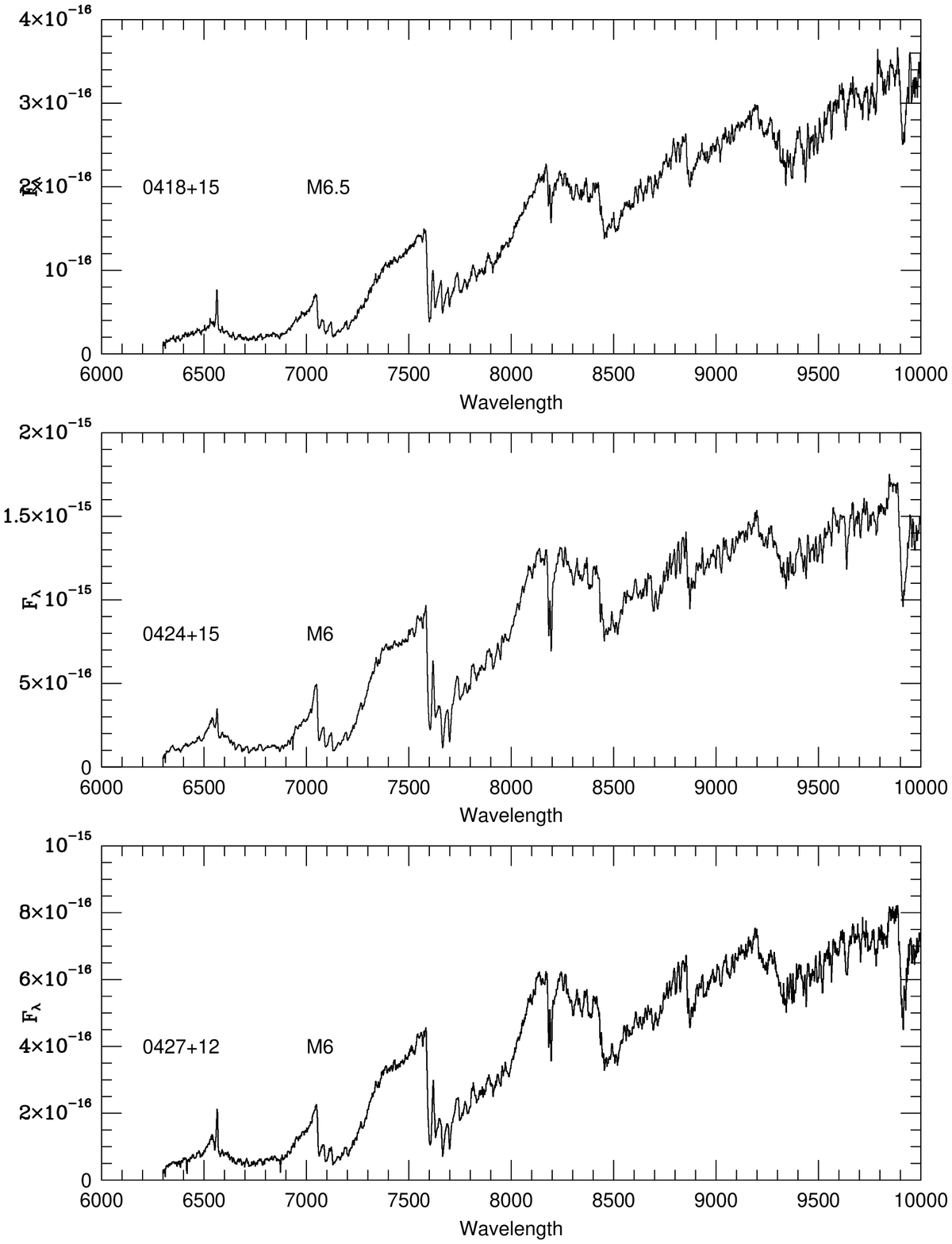}
\caption{  LRIS spectra at spectra at 1.9 \AA pix$^{-1}$}
\end{figure}

\begin{figure}
\plotone{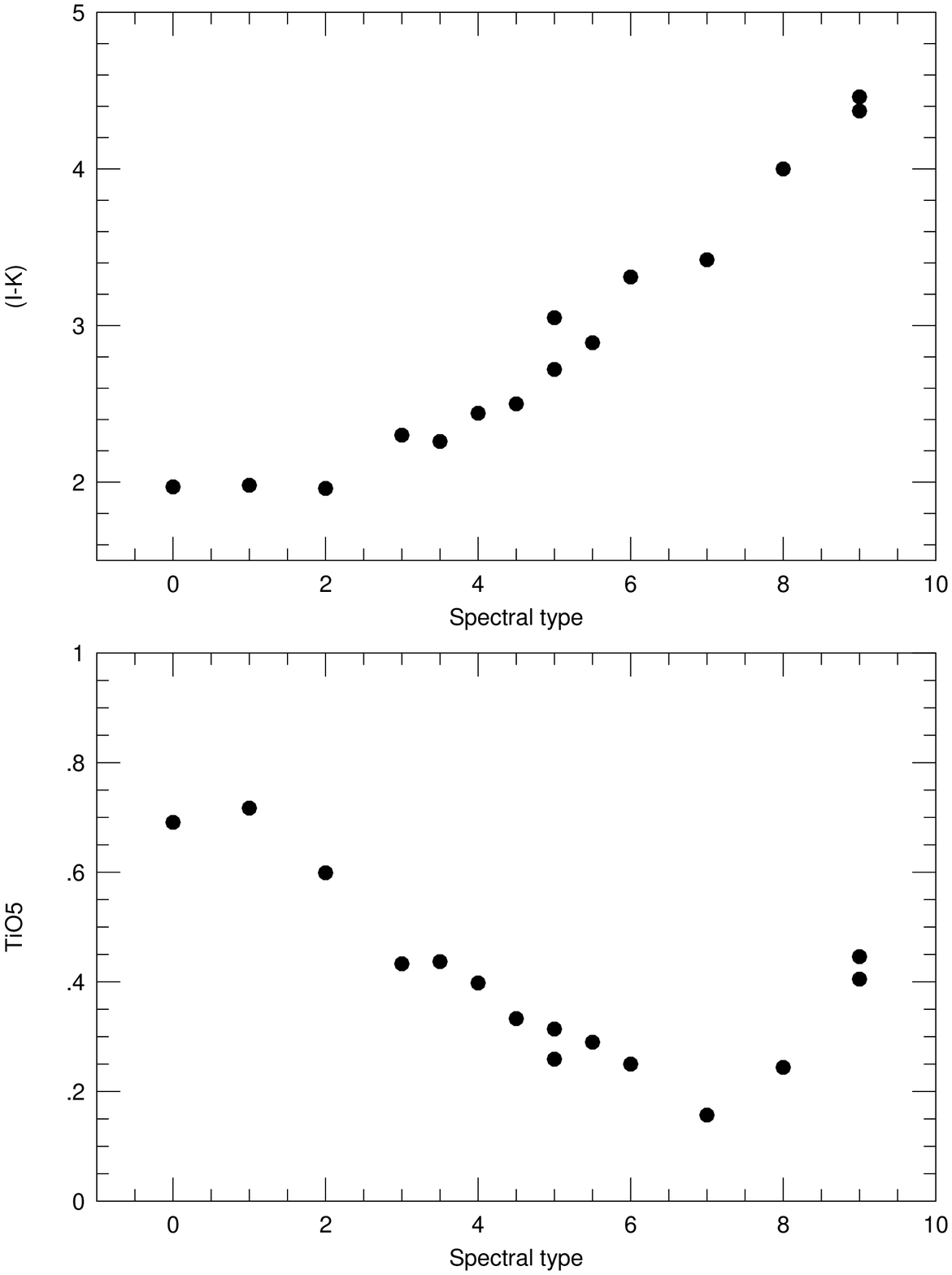}
\caption{ Spectral type calibrations for TiO5 and (I-K)}
\end{figure}

\begin{figure}
\plotone{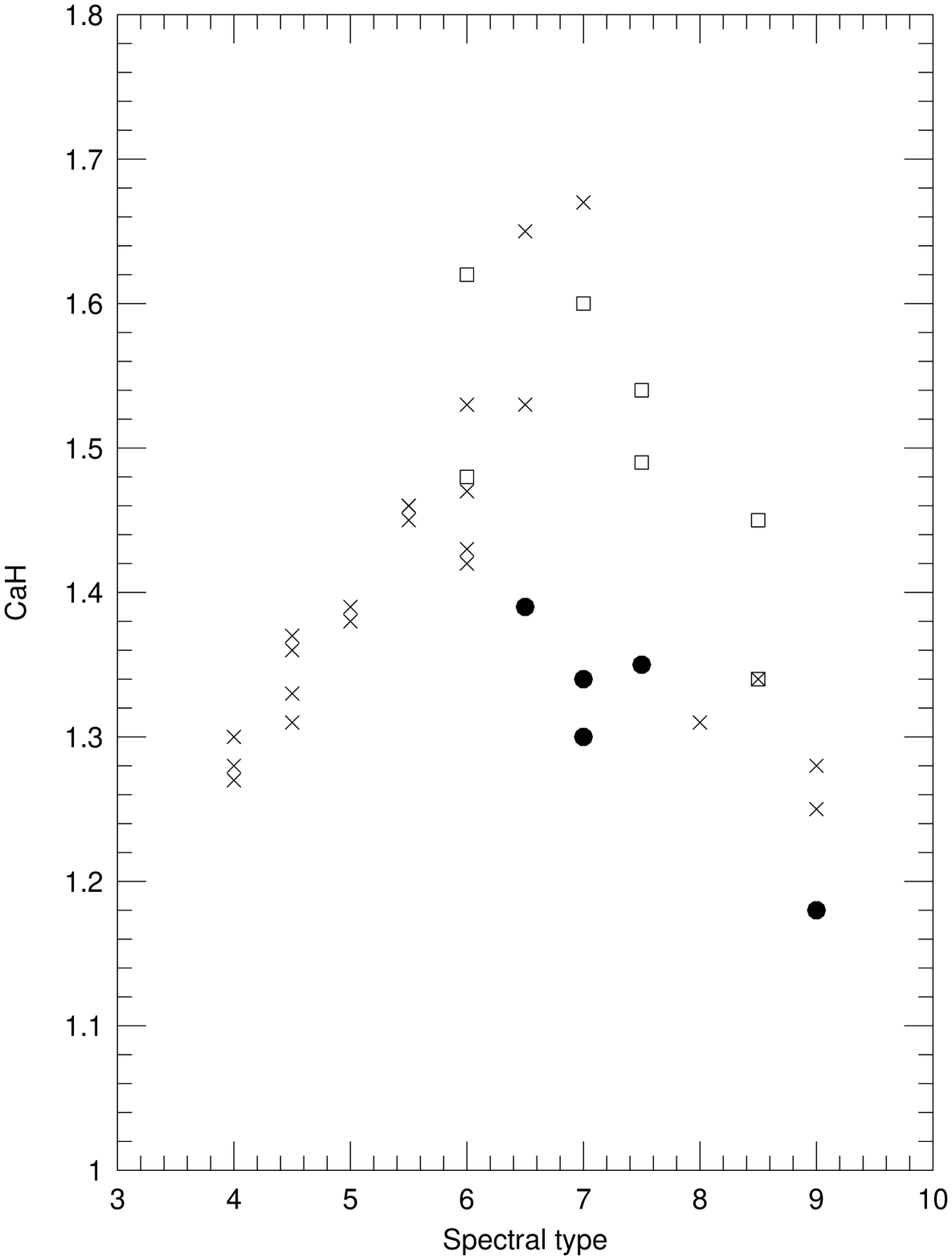}
\caption{ CaH bandstrengths: crosses are spectral standards; open squares
are KI-strong and the solid point KI-weak LH stars}
\end{figure}

\begin{figure}
\plotone{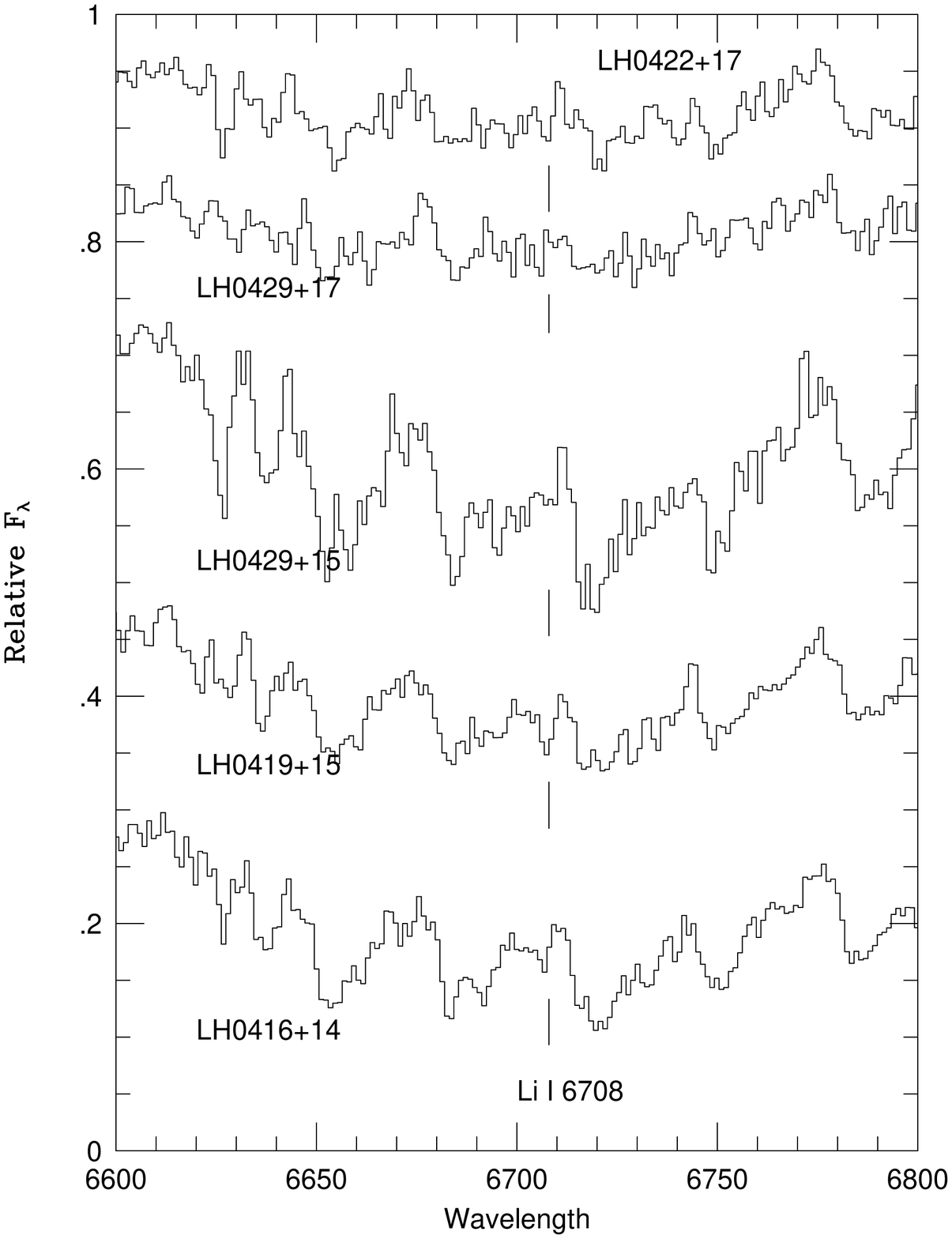}
\caption{ The Li $\lambda 6708 \AA$ region in four KI-weak dwarfs}
\end{figure}

\begin{figure}
\plotone{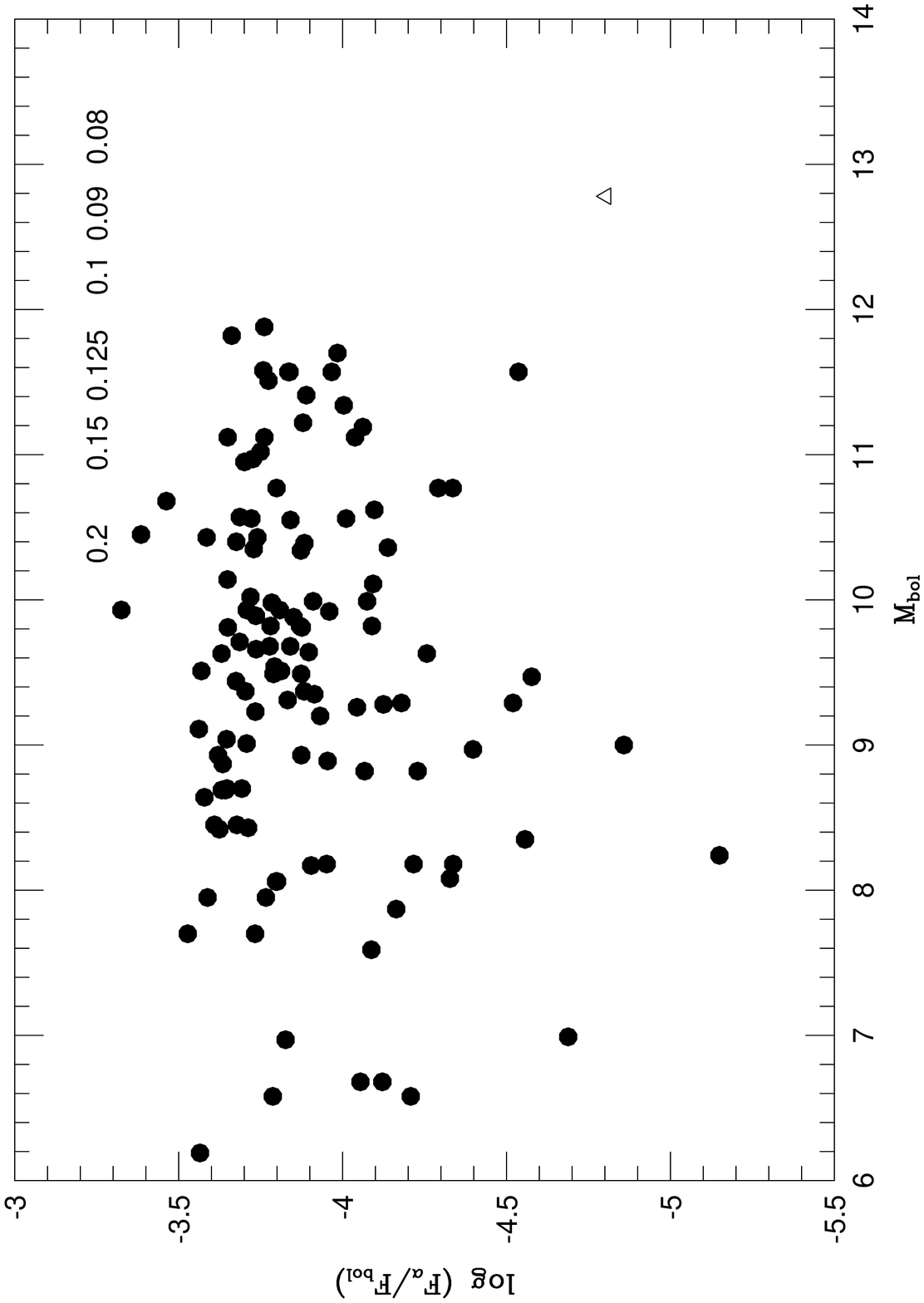}
\caption{ H$\alpha$ activity ratios for Hyades stars. The triangle marks the
position of LH0418+13, if a member.}
\end{figure}

\begin{figure}
\plotone{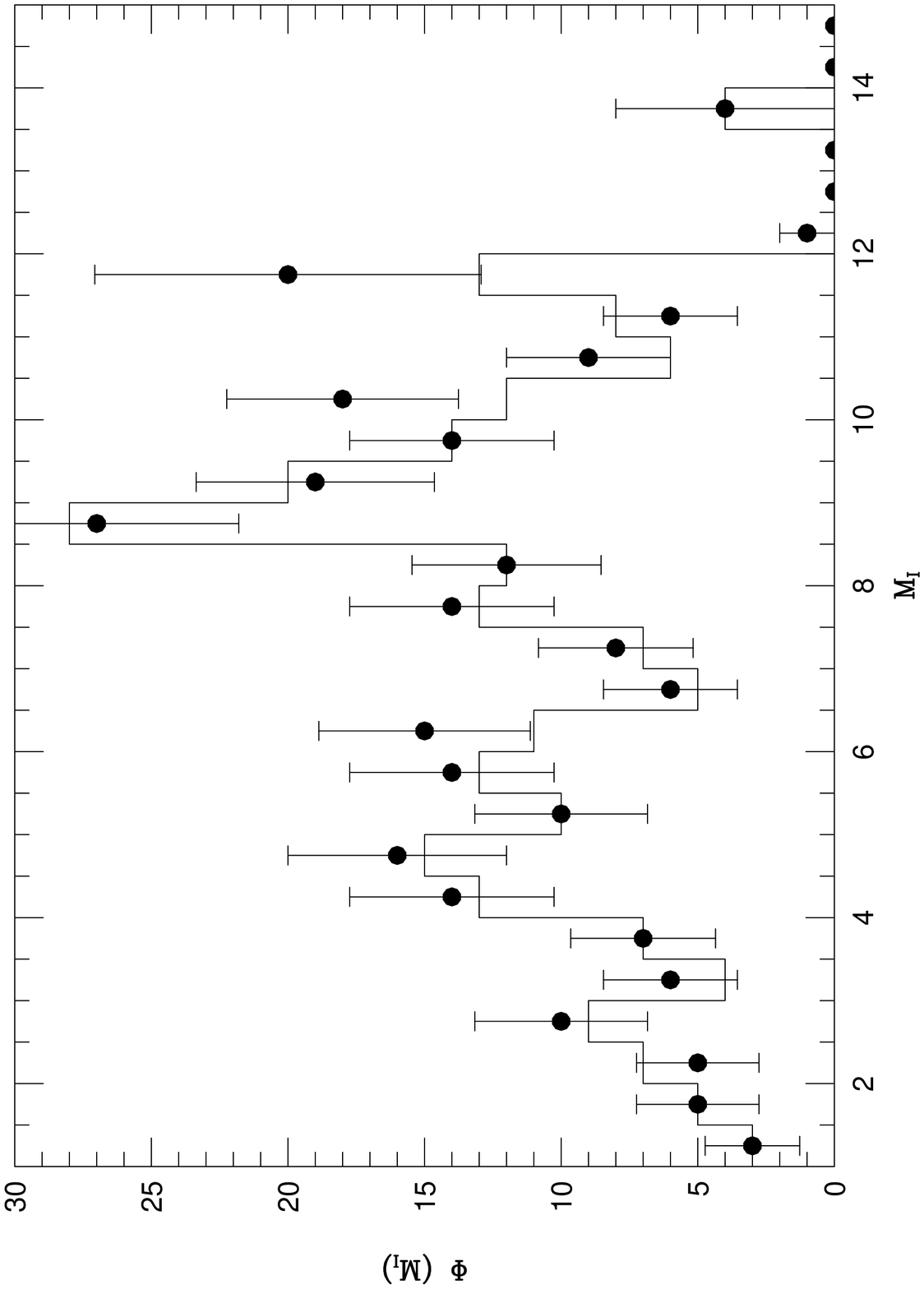}
\caption{ The Hyades I-band luminosity function.}
\end{figure}

\begin{figure}
\plotone{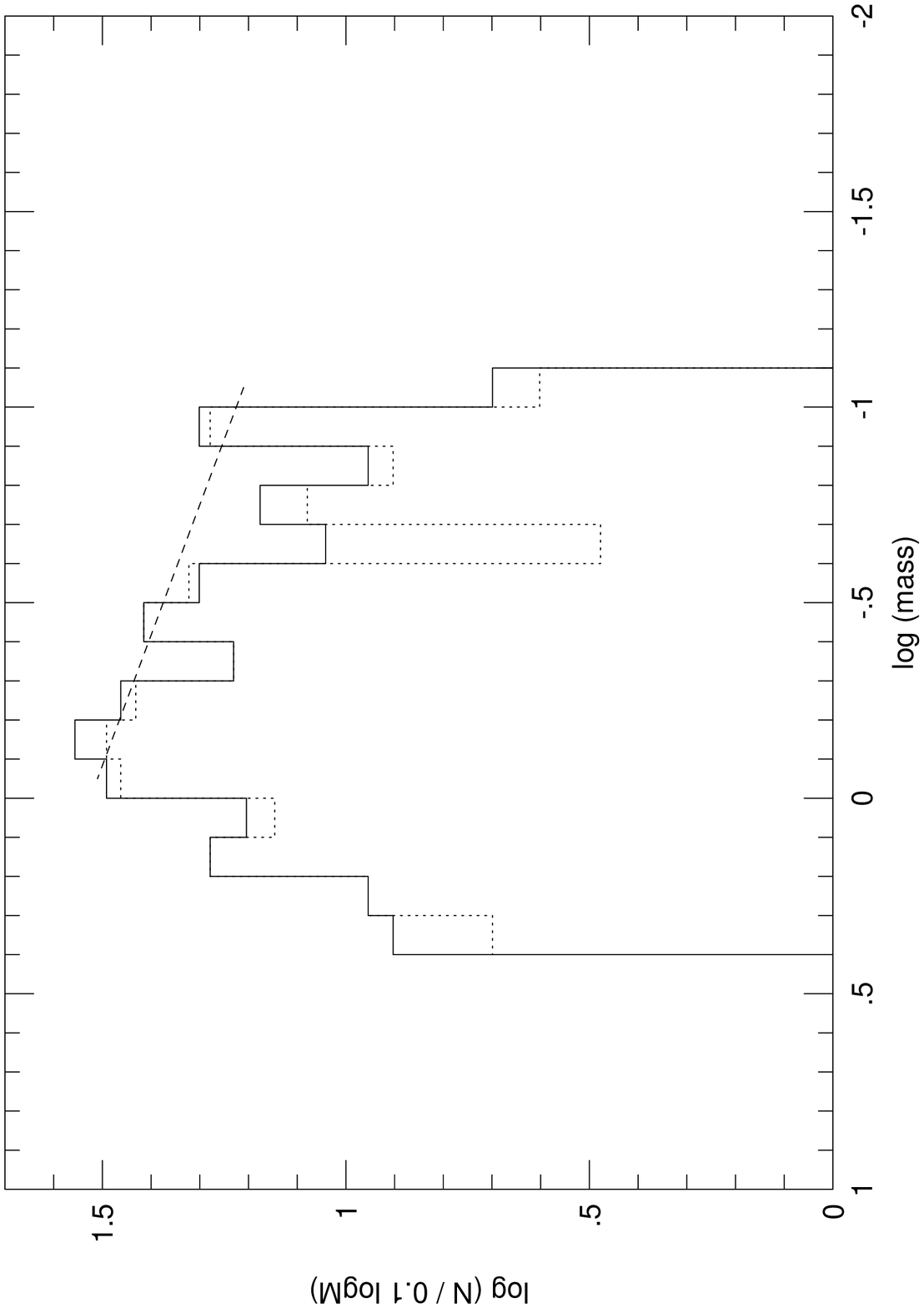}
\caption{ The present-day mass function of the Hyades.}
\end{figure}

\begin{figure}
\plotone{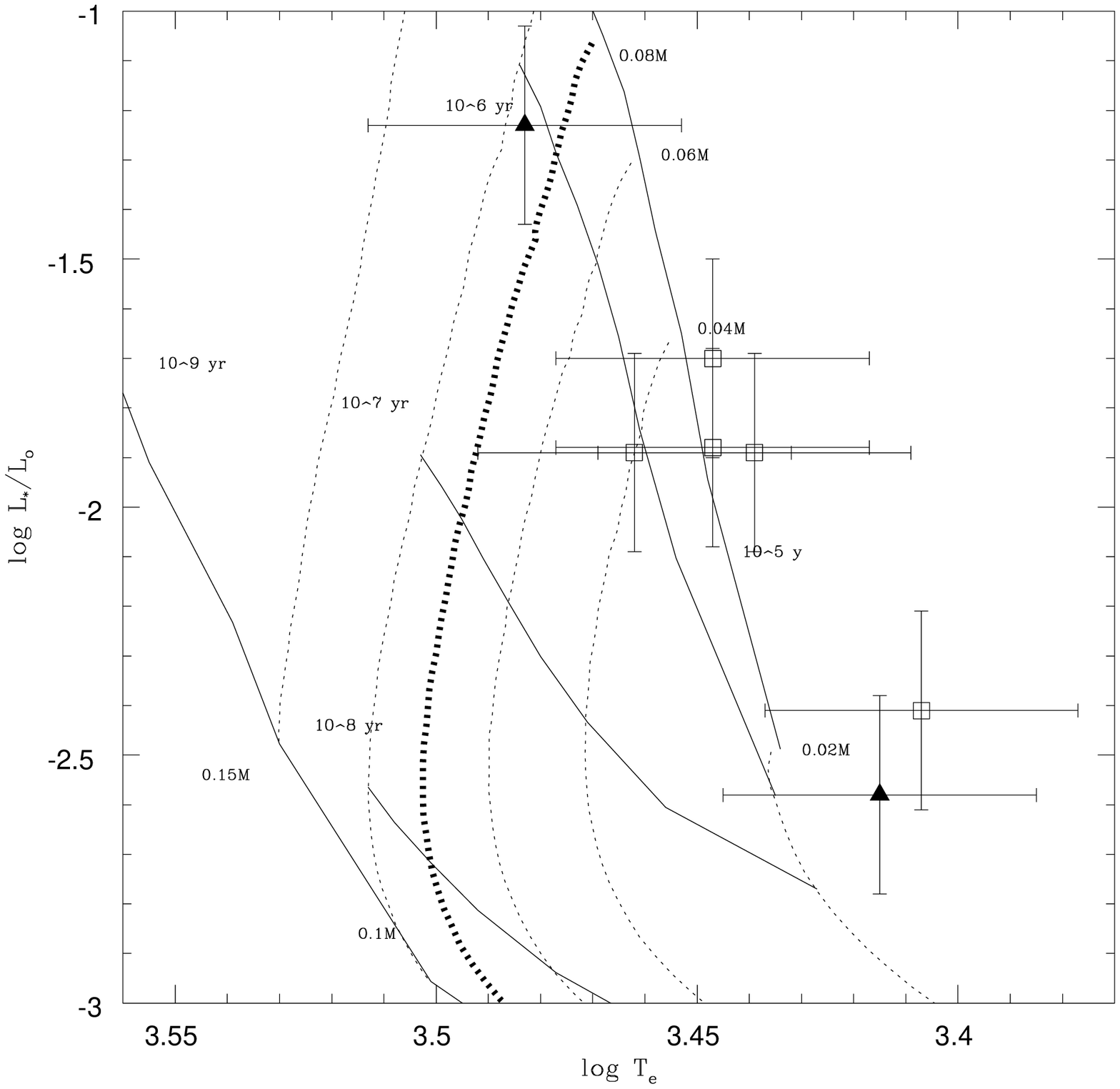}
\caption{ The HR diagram for the LH PMS dwarfs. The tracks are from D'Antona \&
Mazzitelli, 1998.}
\end{figure}

\begin{figure}
\plotone{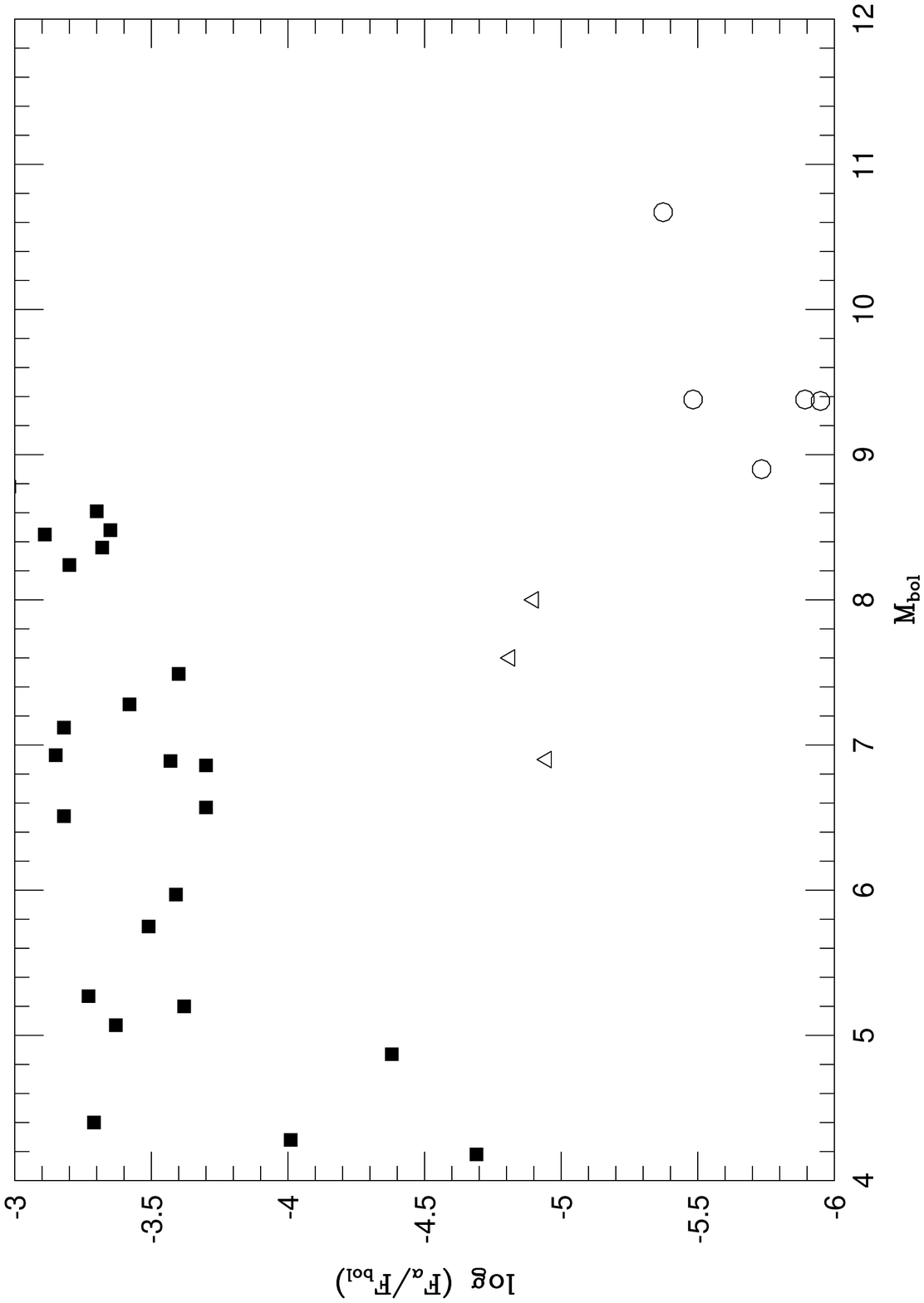}
\caption{ Chromospheric activity in pre-main sequence stars.}
\end{figure}


\begin{thebibliography}{}
\bibitem[]{a1} Bessell, M.S 1986, PASP, 98, 1303
\bibitem[]{a2} Bessell, M.S., Brett, J.M., Scholz, M., Wood, P.R. 1989, AAS, 77, 1
\bibitem[]{}Bonsack, W.K., Greenstein, J.L. 1960, \apj, 131, 83
\bibitem[]{a3} Bryja, C., Humphreys, R.M., Jones, T.J. 1994, AJ, 107, 246
\bibitem[]{a4} Burrows, A., Hubbard, W.B., Saumon, D., Lunine, J.I. 1993, ApJ, 406, 158
\bibitem[]{a5} Burrows, A., Marley, M., Hubbard, W.B., Lunine, J.I., Guillot, T., Saumon, D.,
Freedman, R., Sudarsky, D., Sharp, C. 1997, ApJ, 491, 856
\bibitem[]{a6} D'Antona, F., Mazzitelli, I. 1998, priv. comm.
\bibitem[]{a6a} de la Fuente Marcos, R. 1995, \aap, 301, 407
\bibitem[]{a6b} Delfosse, X., Forveille, T., Perrier, C., Mayor, M. 1998, \aap, 331, 581
\bibitem[]{}Gizis, J.E. 1997, \aj, 113, 806
\bibitem[]{a8} Griffin, R.F., Gunn, J.E., Zimmerman, B.A., Griffin, R.E.M. 1988, AJ, 96, 172
\bibitem[]{a9} Hambly, N.C., Hawkins, M.R.S., Jameson, R.F. 1993, AAS, 100, 607
\bibitem[]{a10} Hamuy, M. et al 1994, PASP, 106, 566
\bibitem[]{a10a} Hanson, R.B. 1975, \aj, 80, 379
\bibitem []{} Harris, H.C. et al 1998 AJ, submitted
\bibitem[]{a11} Hawley, S.L., Gizis, J.E., Reid, I.N. 1996, AJ, 112, 2799
\bibitem[]{a12} Henry, T.J., Kirkpatrick, J.D., Simons, D.A. 1995, AJ, 108, 1437
\bibitem[]{a13} Henry, T.J.. McCarthy, D.W. 1993, AJ, 106, 773
\bibitem[]{a14} Kirkpatrick, J.D., Henry, T.J., McCarthy, D.W.Jr. 1991, ApJS, 77, 417
\bibitem[]{a15} Kirkpatrick, J.D., Kelly, D.M., Rieke, G.H., Liebert, J., Allard,  Wehrse, R. 
1993, ApJ, 402, 643
\bibitem[]{a15} Kirkpatrick, J.D., Henry, T.J., Simons, D.A. 1995, AJ, 109, 797
\bibitem[]{a16} Leggett, S.K. 1992, ApJS, 82, 351
\bibitem[]{a17} Leggett, S.K., Hawkins, M.R.S 1988, MNRAS, 234, 1065
\bibitem[]{a18} Leggett, S.K., Hawkins, M.R.S 1989, MNRAS, 238, 145
\bibitem[]{a19} Leggett, S.K., Harris, H.C., Dahn, C.C. 1994, AJ, 108, 944
\bibitem[]{a20} Leggett, S.K., Allard, F., Berriman, G., Dahn, C.C., Hauschildt, P. 1996, ApJS, 104, 117
\bibitem[]{a21} Luhman, K.L., Liebert, J., Rieke, G.H. 1997, ApJ, 489, L165
\bibitem[]{a22} Luhman, K.L., Briceno, C., Rieke, G.H., Hartmann, L. 1998, ApJ, 493, 909
\bibitem[]{a23} Luyten, W.J, Hill, G., Morris, S., 1981, {\sl Proper Motion Surveys 
with the 48-inch Schmidt telescope}, LIX, University of Minnesota
\bibitem[]{a24} Magazzu, A., Martin, E.L., Rebolo, R. 1993, ApJ, 404, L17
\bibitem[]{a25} Martin, E.L., Rebolo, R., Zapatero-Osorio, M.R. 1996, ApJ, 469, 706
\bibitem[]{mou} Mould, J.R. 1976, \apj 207, 535
\bibitem[]{a26} Neuhauser, R., Torres, G., Sterzik, M.F., Randich, S. 1997, AA, 325, 647
\bibitem[]{a27} Oke, J.B., Gunn, J.E. 1983, ApJ, 266, 713
\bibitem[]{a28} Oke, J. B., Cohen, J. G., Carr, M., Cromer, J., Dingizian, A., 
Harris, F. H., Labreque, S., Lucinio, R., Schaal, W., Epps, H., 
Miller, J. 1995, \pasp, 107, 375
\bibitem[]{a29} Oppenheimer, B.R., Basri, G., Nakajima, T., Kulkarni, S.R. 1998, AJ, 113, 296
\bibitem[]{a30} Patience, J., Ghez, A., Reidn, I.N., Weinberger, A.J., Matthews, K. 1998, ApJ, in press
\bibitem[]{a32a} Pels, G., Oort, J.H., Pels-Kluyvert, H.A. 1975, \aap, 43, 423
\bibitem[]{a31a} Pinsonneault, M.H., Stauffer, J., Soderblom, D.R., King, J.R., 
Hanson, R.B. 1998, \apj, 504, 170
\bibitem[]{a31} Perryman, M.A.C {\sl et al} 1998, A\&A, 331, 81
\bibitem[]{a32} Raboud, D., Mermilliod, J.-C. 1998,AA, in press
\bibitem[]{a33} Reid, I.N., 1992, MNRAS, 257, 257
\bibitem[]{a34} Reid, I.N., 1993, MNRAS, 265, 785
\bibitem[]{a34a} Reid, I.N., Gilmore, G.F. 1982, \mnras, 201, 73
\bibitem[]{a36} Reid, I.N., Gizis, J.E. 1997a, AJ, 113, 2246
\bibitem[]{a35} Reid, I.N., Gizis, J.E. 1997b, AJ, 114, 1992
\bibitem[]{a37} Reid, I.N., Hawley, S.L., Mateo, M. 1995, MNRAS, 272, 828 (RHM)
\bibitem[]{a38} Reid, I.N., Hawley, S.L., Gizis, J.E. 1995, AJ, 110, 1838
\bibitem[]{a39} Schwan, H., 1990, A\&A, 228, 69
\bibitem[]{a40} Stauffer, J.R., Liebert, J., Giampapa, M., Macintosh, B., Reid, N., 
Hamilton, D. 1994, AJ, 108, 160
\bibitem[]{a41} Stauffer, J.R., Liebert, J., Giampapa, M. 1995, AJ, 109, 298
\bibitem[]{a43} Tinney, C.G., Mould, J.R., Reid, I.N. 1993, AJ, 105, 1045
\bibitem[]{a44} Ungerechts, H., Thaddeus, P. 1987, ApJS, 63, 645
\bibitem[]{a43a} van Altena, W.F. 1969, \aj, 74, 2
\bibitem[]{s44a} van Buren, H.G. 1952, BAN, 11, 385
\bibitem[]{a45} Vogt et al, 1994, S.P.I.E., 2198, 362
\bibitem[]{a29a} Zapatero-Osorio, M.R., Rebolo, R., Martin, E.L., Basri, G.,
magazzu, A., Hodgkin, S.T., Jameson, R.F., Cossburn, M.R. 1997, \apj, 491, L81

\end{thebibliography}
\end{document}